\documentclass{aa}
\usepackage{graphicx}
\usepackage{natbib}
\usepackage{amsmath}
\usepackage{amsfonts}
\bibpunct{(}{)}{;}{a}{}{,}

\newcommand{\Ab}{\boldsymbol{A}}
\newcommand{\Bb}{\boldsymbol{B}}
\newcommand{\Cb}{\boldsymbol{C}}

\newcommand{\xb}{\boldsymbol{x}}
\newcommand{\zb}{\boldsymbol{z}}
\newcommand{\yb}{\boldsymbol{y}}

\newcommand{\eb}{\boldsymbol{e}}
\newcommand{\fb}{\boldsymbol{f}}
\newcommand{\ftb}{\boldsymbol{\tilde f}}

\newcommand{\hb}{\boldsymbol{h}}
\newcommand{\Ib}{\boldsymbol{I}}
\newcommand{\Hb}{\boldsymbol{H}}
\newcommand{\Kb}{\boldsymbol{K}}

\newcommand{\ssb}{\boldsymbol{s}}
\newcommand{\PPb}{\boldsymbol{P}}
\newcommand{\Rb}{\boldsymbol{R}}
\newcommand{\Sb}{\boldsymbol{S}}
\newcommand{\wb}{\boldsymbol{w}}
\newcommand{\uub}{\boldsymbol{u}}
\newcommand{\Ym}{\boldsymbol{\mathcal{Y}}}

\newcommand{\RB}{\mathbb{R}}
\newcommand{\thetawb}{\boldsymbol{\widehat \theta}}

\newcommand{\xhb}{\boldsymbol{\widehat x}}
\newcommand{\yhb}{\boldsymbol{\widehat y}}
\newcommand{\xdb}{\boldsymbol{\dot x}}
\newcommand{\zdb}{\boldsymbol{\dot z}}
\newcommand{\alphab}{\boldsymbol{\alpha}}

\newcommand{\epsilonb}{\boldsymbol{\epsilon}}
\newcommand{\thetab}{\boldsymbol{\theta}}
\newcommand{\sigmab}{\boldsymbol{\sigma}}
\newcommand{\Psib}{\boldsymbol{\Psi}}
\newcommand{\sigmatb}{\boldsymbol{\tilde \sigma}}
\newcommand{\Sigmab}{\boldsymbol{\Sigma}}
\newcommand{\Phib}{\boldsymbol{\Phi}}

\begin{document}

   \title{Time series analysis in Astronomy: \\ limits and potentialities}

   \author{R. Vio\inst{1}
          \and
          N.R. Kristensen\inst{2}
          \and
          H. Madsen\inst{3}
          \and
          W. Wamsteker\inst{4}
          }
   \institute{Chip Computers Consulting s.r.l., Viale Don L.~Sturzo 82,
              S.Liberale di Marcon, 30020 Venice, Italy\\
              ESA-VILSPA, Apartado 50727, 28080 Madrid, Spain\\
              \email{robertovio@tin.it}
         \and
             Department of Informatics and Mathematical Modelling, Technical University of Denmark, 
             Richard Petersens Plads, DK-2800 Kgs. Lyngby, Denmark \\
             \email{nkr@imm.dtu.dk}
         \and
             Department of Informatics and Mathematical Modelling, Technical University of Denmark, 
             Richard Petersens Plads, DK-2800 Kgs. Lyngby, Denmark \\
             \email{hm@imm.dtu.dk}
         \and
             ESA-VILSPA, Apartado 50727, 28080 Madrid, Spain\\
             \email{willem.wamsteker@esa.int}
             }

\date{Received .............; accepted ................}

\abstract{In this paper we consider the problem of the limits concerning the physical information that can be extracted from the analysis of one or more time series (light curves) typical of astrophysical objects. On the basis of theoretical considerations and numerical simulations, we show that with no a priori physical model
there are not so many possibilities to obtain interpretable results. For this reason, the practice to develop more and more sophisticated statistical methods of time series analysis is not very productive. Only techniques of data analysis developed in a specific physical context can be expected to provide useful results. The field of {\it stochastic dynamics}
appears to be an useful framework for such an approach. In particular, it is shown that modelling the experimental time 
series by means of the stochastic differential equations (SDE) represents a valuable tool of analysis. For example,
the use of SDE permits to make the analysis of a continuous signal independent from the frequency sampling
with which the experimental time series have been obtained. In this
respect, an efficient approach based on the extended Kalman-filter technique is presented. Freely downloadable
software is made available.
}
\titlerunning{Modelling of astronomical time series}
\authorrunning{R. Vio et al.}
\maketitle

\section{Introduction}

The study of the light curves of astrophysical objects has always been an important tool for astronomers. The reason is simple: an effective way to get insight on the structure of a given physical system is to study its evolution over time. Some examples are the reconstruction of the structure of the binary star systems, the understanding of the nature of the pulsars, and the determination of the sizes of the central regions of the active galactic nuclei. However, 
in spite of these remarkable successes, in many other situations the analysis of the light curves has not proved to be 
so useful. The reason can be understood by taking into account that often the time evolution of a 
dynamical system, describable in terms of a set of generic physical quantities ({\it state-variables}) $\xb(t)$
\footnote{Hereafter, vector quantities will be denoted in boldface.}, is governed by a n-dimensional system of
differential equations ({\it state-equation}) with the general form
\begin{equation} \label{eq:evolves}
\xdb(t) = \fb[\xb(t), \ssb(t), t],
\end{equation}
where $t$ is the time coordinate, symbol ``~$\boldsymbol{\dot{}}~$'' means derivation with respect to $t$, 
and $\fb[\cdot]$ is a $n$-dimensional (possibly non-linear) function. 
The $m$-dimensional vector $\ssb(t)$ represents 
independent processes whose time evolution 
does not depend on $\xb(t)$ such as, for example, the processes that take place in regions external to the 
system of interest.
In general, the quantities $\xb(t)$ are not directly observable and the experimental time series 
$\{ \yb_{t_k} \}_{k=0}^N$ are obtained through the {\it measurement-equation}
\begin{equation}
\yb_{t_k} = \hb[\xb(t_k), t_k] + \eb_{t_k}.
\end{equation}
Here, $\hb[\cdot]$ is a $l$-dimensional (possibly non-linear) function, $\{ t_k \}_{k=0}^N$ 
is the set of sampling time instants, and $\{ \eb_{t_k} \}_{k=0}^N$ represents 
the measurement errors. Usually, the number $l$ of available time series 
is smaller than $n$, and often $l=1$.
This means that the observed signals provide information only on a projection of the dynamics 
of the system of interest. Curiously, instead of developing new methodologies for the analysis of the observed signals 
in a given physical context, in the past much effort has been spent in an attempt to devise more and more 
sophisticated techniques for the {\it statistical} characterization of $\yb(t_k)$ (e.g. AR and ARMA modelling, 
maximum entropy power spectra, ...). The expectation of these efforts was that such a characterization could be able to provide hints on the functional form of $\fb[.]$. However, the results have very often been disappointing 
since the experimental time series do not contain all the information necessary for such a task. 

This does not mean that the classic statistical analysis of the time series is unproductive. However, it has to represent only a starting point, otherwise there is the risk that the studies on the time evolution of the astrophysical
objects could merely consist in a collection of data and in the elaboration of some generic statistical measures with no direct physical meaning.

On the basis of this argument, \citet{vio92} stress the necessity to start directly from model~(\ref{eq:evolves}),
as provided by a given theoretical model, and to use time series as a test for such model. 
More in particular, from the consideration that many astrophysical objects show evolutions that are unpredictable 
over time, they suggest to modify Eq.~(\ref{eq:evolves}) to
\begin{equation} \label{eq:evolves1}
\xdb(t) = \fb[\xb(t), \uub(t), \wb(t), t, \thetab],
\end{equation}
with $\uub(t)$ and $\wb(t)$ representing deterministic and random processes, respectively, 
and then to solve them via a numerical approach. In other words, the functional form of $\fb[ \cdot ]$ 
is assumed known out of a set $\thetab$ of parameters. In this way, once fixed the values of $\thetab$,
it is possible to obtain  ``{\it synthetic}'' 
light curves that can be compared with the observed ones. The reason to add the random process $\wb(t)$,
typically a continuous 
Gaussian white noise\footnote{NB. Hereafter, with the term noise we will indicate only the random 
processes perturbing the dynamics 
of a given physical system and not the contamination due to the measurement errors.}, is a consequence of
the fact that this term represents the interaction of the physical system of interest with its surroundings and/or the action of complex processes that cannot be directly included into the model (e.g. gas turbulence). In general, such processes are characterized by an huge number of degrees of freedom and therefore they can be assumed to have a 
stochastic nature. In practice, this means to study the time evolution of a given physical system in 
the context of the so called {\it stochastic dynamics}, i.e., through the modelling of the observed time series 
by means of stochastic differential equations (SDE). 

Although {\it stochastic dynamics} is an approach widely used for the study and simulation of realistic
scenarios in many fields of applied science and engineering as, for example, fluid dynamics, structural and 
mechanical engineering, avionics, material properties, financial sciences \ldots \citep{gha91, klo97, gar99}, 
in Astronomy it is not so known. However, as we will show on the basis of some numerical experiments, even 
the action of quite weak noise sources on simple nonlinear dynamical systems can produce deep modifications 
of their behaviour over time. This means that in real scenarios the noise component must be considered as intrinsic 
to the physics of the systems and not only a secondary factor. Consequently, in many situations, {\it stochastic
dynamics} could represent the only possibility to use the experimental time series in an effective way. 

In \citet{vio92} it is suggested that the value of $\thetab$ has to be derived from physical considerations.
Here, we provide some tools that permits to estimate $\thetab$ directly from the data.

In Sect.~\ref{sec:why} some arguments are presented that support the necessity of stochastic modelling in the
study of the physical systems, and in Sec.~\ref{sec:astronomy} an example in the astronomical context is provided.
In Sec.~\ref{sec:discrete} it is shown that modelling the time series through discrete models is
unsuited for most of physical systems. Hence, in Sec.~\ref{sec:continuum} an approach based on
SDE is suggested and some tools are provided for using it in modelling the time
series. An example of application is given in Sec.~\ref{sec:worked}. Finally, the conclusions
and some final comments are presented in Sec.~\ref{sec:conclusions}. 

\section{Why is stochastic modelling necessary?} \label{sec:why}

\begin{figure}
        \resizebox{\hsize}{!}{\includegraphics{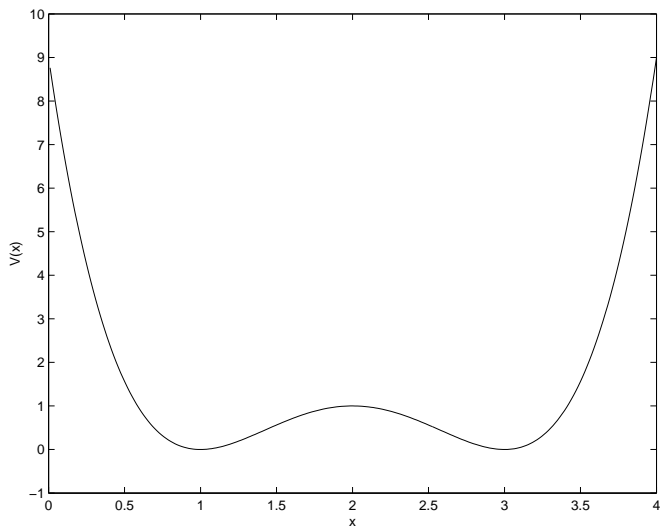}}
        \caption{Potential function associated to the dynamical model (\ref{eq:random}).}
        \label{fig:potential}
\end{figure}

In this section we will consider the time evolution of the simple dynamical system
\begin{equation} \label{eq:random}
\dot x(t) = -x^3(t) + 6 x^2(t) - 11 x(t) + 6
\end{equation}
suffering the influence of different kinds of noise. This is a nonlinear model with the particularity that the 
associated potential is characterized by two regions, both of them with their own stable point of equilibrium 
(see Fig.~\ref{fig:potential}). We have 
deliberately chosen to represent an unsophisticated physical situation since our aim is to show that, even in this simplified scenario, the fact of not considering the noise as a fundamental component of the dynamics makes essentially impossible to get insights on the characteristics of the system under study.

\subsection{Perturbation by an additive noise process}

Figs.~\ref{fig:series}a,b show two realizations of the stochastic differential equation
\begin{equation} \label{eq:random1}
\dot x(t) = -x^3(t) + 6 x^2(t) - 11 x(t) + 6 + \sigma w(t),
\end{equation}
where $w(t)$ is a zero-mean, unit-variance, and continuous white noise process. In both the simulations the same realization of $w(t)$ is used but two different values, respectively, $0.1$ and $0.5$, have been adopted for the constant $\sigma$. Notice that in this example $w(t)$ acts as a simple additive perturbing process.

Clearly, the two signals show a different time evolution. The reason lies in the two points of equilibrium that 
characterize the dynamics of model (\ref{eq:random}). Indeed, the only possibility for the system to jump from one 
region to the other is represented by the perturbation $\sigma w(t)$. If such perturbation is strong enough, then 
the jumps will be frequent and the bi-stable characteristic of the system will be revealed even by short observed 
signals. Conversely, if the perturbation is small, then it is possible that for observing a single jump it is 
necessary to wait for a long time. From the statistical analysis of $x(t)$ it is possible to obtain some results 
also in this unfavourable situation. For example, the Keenan test \citep{kee85}, a test devised for verifying the nonlinearity of the time series, is able to detect the nonlinear nature of the data sequence shown in 
Fig.~\ref{fig:series}a at a confidence level of $95\%$. However, it is superfluous to stress that, if no jump is 
detected, even the most sophisticated statistical analyses will be unable to provide more detailed information on 
the functional form of model (\ref{eq:random}).

\subsection{Perturbation by a multiplicative noise process}

\begin{figure}
        \resizebox{\hsize}{!}{\includegraphics{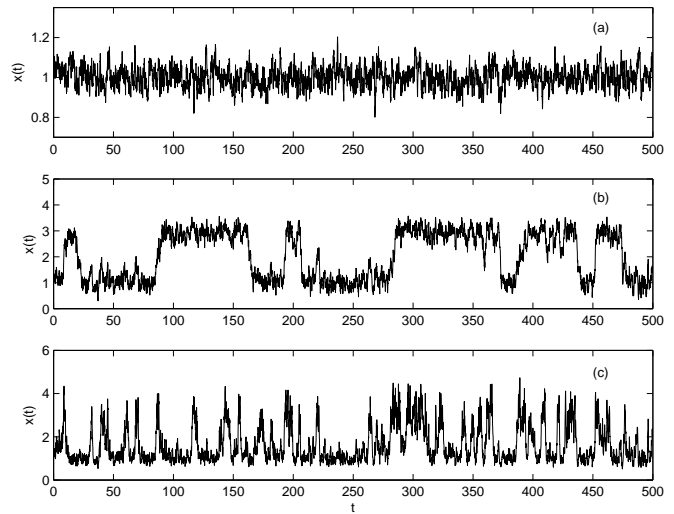}}
        \caption{a), b) Time series obtained from the dynamical model (\ref{eq:random1}) with $\sigma$ equal, 
	  respectively, to $0.1$ and $0.5$; c) Time series obtained from the dynamical model (\ref{eq:random2}) 
	  with $\sigma$ equal to $0.5$.}
        \label{fig:series}
\end{figure}
\begin{figure}
        \resizebox{\hsize}{!}{\includegraphics{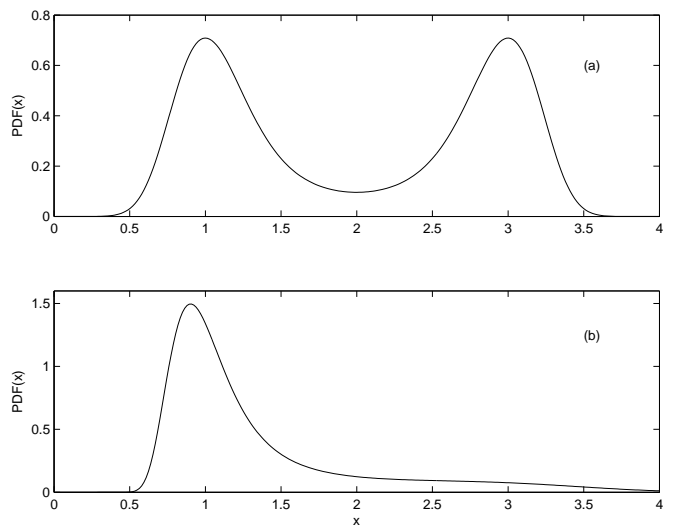}}
        \caption{Probability density functions associated a) to the dynamical model (\ref{eq:random1}) 
	  and b) to the dynamical model (\ref{eq:random2}).}
        \label{fig:distributions}
\end{figure}

The situation worsens when $x(t)$ affects the intensity of the perturbation process. This is shown by 
Fig.~\ref{fig:series}c that presents a realization of the stochastic system
\begin{equation} \label{eq:random2}
\dot x(t) = -x^3(t) + 6 x^2(t) - 11 x(t) + 6 + \sigma x(t) w(t).
\end{equation}
Here, the realization of process $w(t)$ is the same used in the previous example, and the value of $\sigma$ is equal 
to that used in Fig.~\ref{fig:series}b. With respect to model~(\ref{eq:random1}), now the standard deviation of the 
noise component at time $t$, being given by $\sigma x(t)$, is not constant but depends on the value of the signal at 
the same time instant. 

A comparison of Fig.~\ref{fig:series}b with Fig.~\ref{fig:series}c indicates that, although the only modification regards the noise component,  the dynamical behaviour of $x(t)$ has suffered deep changes. This is illustrated in Figs.~\ref{fig:distributions}a,b which show the probablity density functions (PDF) associated to the two processes. From Fig.~\ref{fig:distributions}b it is evident that the bi-stable structure of model (\ref{eq:random}) is no more detectable. The problem is that signal $x(t)$, because of the term $x(t) w(t)$, is no longer able to furnish direct information on the deterministic part of system (\ref{eq:random2}). More than in the previous example, this illustrates that an approach based only on the statistical analysis of the experimental data can be quite non-informative, no matter how sophisticated the technique applied.

Unfortunately, it is highly probable that a situation like this one constitutes a typical situation for many astrophysical objects. For example, it is to be expected that the luminosity of a given object could depend on the quantity of gas accreting onto it, while, at the same time, it is conceivable that the accretion rate is influenced by the energy emitted by the object itself.

\section{An astronomical example} \label{sec:astronomy}

In order to show that neglecting the stochastic component of a nonlinear dynamical system can be risky also in case
of astrophysical systems, here we give an example based on the {\it pair-production instability model} by 
\citet{mos86}, that has been proposed to explain the strongly variable emission of high-energy radiation from the 
central regions of the active galactic nuclei (AGN). Since we want to maintain the readability of the paper also 
outside the context of the physics of AGN, the details of this model will be not given.

According to this model, the strongly variable emission in the hard X-ray electromagnetic waveband of AGN's can be modelled via a scenario where the gas, accreting a black hole residing in the central regions of these objects, suffers  a ``pair production instability'' (an instability due to the creation of electron-positron pairs). The time evolution of such a system can be formalized via the following set of differential equations describing, respectively, 1) pair creation and annihilation; 2) photon production, absorption and escape; 3) electron/positron heating and cooling; and 4) the proton density changes:
\begin{eqnarray}
\frac{d n_+}{d t} &=& \dot n_+^{{\rm cre}} - \dot n_+^{{\rm ann}} \label{eq:accretion1}\\
\frac{d n_{\gamma}}{d t} &=& \frac{\dot u_{{\rm cbr}}}{k T_e} - \frac{3 n_j}{1+\tau_T} \frac{c}{R}
-2 \frac{d n_+}{d t} \label{eq:accretion2}\\
\frac{d u_e}{d t} &=& \dot u_{{\rm ep}} - \dot u_{{\rm cbr}} \label{eq:accretion3}\\
\frac{d n_p}{d t} &=& \lambda_0 - \frac{n_p}{t_{{\rm ep}}} \label{eq:accretion4}
\end{eqnarray} 
where $u_e \simeq n_e k T_e$, $T_e=$ electron temperature, $n_e$, $n_+$, $n_{\gamma}$, and $n_p$ are, respectively, the densities of the electrons, positrons, photons, and protons, $\tau_T = n_e R \sigma_T$, $\sigma_T =$ Thomson cross-section, $\dot n_+^{{\rm cre}}=$ pair creation rate, $\dot n_+^{{\rm ann}}=$ pair annihilation rate, $\dot u_{{\rm ep}}=$
rate of energy transfer from protons to electrons, $\dot u_{{\rm cbr}}=$ electron cooling rate, $\lambda_0=$ quantity proportional to the accretion rate of the gas surrounding the central regions of AGN's , and  $t_{{\rm ep}}=$ electron energy transfer time. In the present context, the observed X-ray light curves can be assumed to be proportional to the quantity $n_{\gamma}(t)$.

\begin{figure}
        \resizebox{\hsize}{!}{\includegraphics{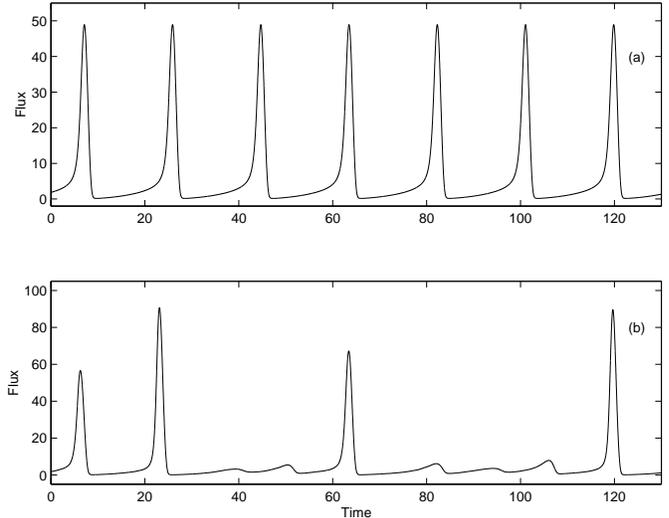}}
        \caption{a) Light curve $n_{\gamma}(t)$ obtained by the numerical integration of the system of equations 
        (\ref{eq:accretion1})-(\ref{eq:accretion4}); b) The same light curve when Eq.~(\ref{eq:accretion4}) 
        is substituted with Eq.~(\ref{eq:accretion5}).}
        \label{fig:pair}
\end{figure}

An interesting point is that, for certain values of the parameters, this system gives rise to periodic hard X-ray 
flares. In the recent past, this model has enjoyed a certain fame because of this ability. However, a serious 
drawback of the above scenario is that the accretion rate of the gas is supposed to be strictly constant. Of course, 
this is a strong assumption and it raises some doubts on the reliability of the periodic behaviour under 
realistic conditions. 

Figs.~\ref{fig:pair}a,b show what happens to a periodic light curve, obtained from the numerical solution of 
Eqs.~(\ref{eq:accretion1})-(\ref{eq:accretion4}), when the accretion rate $\lambda_0$ is perturbed by a continuous, additive white noise process with standard deviation $\sigma = 0.15 ~\lambda_0$. Formally, this means to substitute 
Eq.~(\ref{eq:accretion4}) with:
\begin{equation} \label{eq:accretion5}
\frac{d n_p}{d t} = [\lambda_0 +\sigma w(t)] - \frac{n_p}{t_{{\rm ep}}},
\end{equation}
that certainly represents a much more realistic assumption.
It is evident that, although the perturbation is not so strong, the periodic behaviour of the light curve has 
almost completely disappeared.

The indication provided by this experiment is that the modelling of astrophysical systems via deterministic equations 
can be misleading. Indeed, some features can be forecasted but, probably, are very difficult to be actually observed 
in experimental data.

\section{Is discrete modelling appropriate?} \label{sec:discrete}

In the previous sections it has been shown that neglecting the noise component in the dynamics of the astrophysical
systems can be risky as concerns their expected time evolution. Moreover, a simple statistical approach
appears inadequate. Hence, some modelling is necessary. In this respect, since the experimental
time series are discrete in nature, one could be tempted 
to adopt a discrete approach for modelling the observed signals. Actually, this is not so appropriate 
as it could appear at first sight.
In fact, if the continuous nature of the signals is not taken into account, the results provided by any method of
analysis are in general to be expected to depend on the sampling time. 

In order to show this point it is useful to consider a simple stochastic model:
\begin{equation} \label{eq:first}
\dot x = - \theta x(t) + \sigma w(t); ~\qquad \theta \ge 0,
\end{equation}
where $\theta$ and $\sigma$ are constants.
It is not difficult to see that, when this system is observed at a set of discrete, evenly spaced time instants,
its dynamics can be described in terms of the discrete model
\begin{equation} \label{eq:discrete}
x_{k+1} = \alpha x_k + w_{k},
\end{equation}
where $x_k = x(t_k)$, and $\{ w_k \}_{k=0}^{N}$ is the realization of a discrete white noise process. From the point 
of view of classical time
series analysis, Eq.~(\ref{eq:discrete}) represents an AR($1$) model. This fact could suggest the possibility to
obtain information
on system (\ref{eq:first}) by means of the classical discrete AutoRegressive modelling. Unfortunately, this is
not true because the relationship between $\theta$ and $\alpha$ \citep{vio92},
\begin{equation}
\alpha = \exp(-\theta \Delta t),
\end{equation}
depends on the sampling time step $\Delta t$. The consequence is that, when the sampling frequency decreases, the
values of $\alpha$ goes to zero. In other words, if signal $x(t)$ is observed at discrete instants, it will tend
to appear as a white noise for $\Delta t \rightarrow \infty$. 

Another problem comes out from the fact that the relationship between the parameters of the continuous and 
of the discrete models can be complex and difficult to recover.
For example, the second order linear system,
\begin{equation}
\ddot x(t)+ \theta_1 \dot x(t) + \theta_0 x(t) = w(t),
\end{equation}
can be shown \citep{pan75} to be equivalent to a discrete ARMA($2$,$1$) model,
\begin{equation} \label{eq:arma}
x_{k+1} - \phi x_k - \phi_2 x_{k-1} = w_{k} - \psi w_{k-1},
\end{equation}
where,
\begin{align}
\phi_1 & = \exp(\mu_1 \Delta t) + \exp(\mu_2 \Delta t); \\
\phi_2 & = \exp[(\mu_1 + \mu_2) \Delta t];
\end{align}
\begin{equation}
\psi = \frac{\psi_1}{\psi_2};
\end{equation}
\begin{align}
\psi_1 = & (\mu_1 - \mu_2) [ \mu_2 {\rm e}^{\mu_2 \Delta t}  - \mu_1 {\rm e}^{\mu_1 \Delta t} \nonumber \\
& + {\rm e}^{[(mu_1 + \mu_2) \Delta t]} (\mu_2 {\rm e}^{\mu_1 \Delta t}  - \mu_1 {\rm e}^{\mu_2 \Delta t})];
\end{align}
\begin{align}
\psi_2 = & (\mu_2 {\rm e}^{\mu_1 \Delta t} - \mu_1 {\rm e}^{\mu_2 \Delta t})^2 \nonumber \\
& + \mu_1 \mu_2 ({\rm e}^{\mu_1 \Delta t} - {\rm e}^{\mu_2 \Delta t})^2 - (\mu_1 - \mu_2)^2;
\end{align}
\begin{equation}
\mu_1, \mu_2 = \frac{-\theta_1 \pm \sqrt{\theta_1^2 - 4 \theta_0}}{2}.
\end{equation}
Apart from the above mentioned dependence on $\Delta t$, it is important to note that, differently from the classical
ARMA models, model~(\ref{eq:arma}) has only two independent parameters: once $\mu_1$ and $\mu_2$ are known,
the three parameters $\phi_1$, $\phi_2$ and $\psi$ are fixed. This fact implies that the classical ARMA($2$,$1$) model
is inadequate to represent model~(\ref{eq:arma}) because it assumes the independence of the characteristic
parameters.

The situation becomes critical in case of nonlinear systems since, in general, it is even not possible to
infer the relationship between the parameters of the continuous models and those of the discrete ones.
The reason is easy to understand by considering the following one-dimensional model
\begin{equation} \label{eq:diffusion}
\dot x(t) = \mu[x(t), \thetab] + \sigma[x(t), \thetab] w(t).
\end{equation}
If the sampling frequency is sufficiently high, Eq.~(\ref{eq:diffusion}) can be approximated by
\begin{equation} \label{eq:approx}
x_{k+1} = x_k + \mu[x_k, \thetab] \Delta t + w_k \sigma[x_k, \thetab] (\Delta t)^{1/2},
\end{equation}
i.e., with a discrete model,
\begin{equation}
x_{k+1} = h(x_k) + \epsilon_k,
\end{equation}
where $h(x_k)$ is a function that can be related back to the parameter vector $\thetab$,
and $\{ \epsilon_k \}$ are independent, discrete Gaussian random quantities. The important point is that, 
with this model, parameters $\thetab$ can be estimated through a classical maximum likelihood method.
Things are more intricate if, as it usually happens in practical applications, $\Delta t$ is not so small 
to make approximation~(\ref{eq:approx})
to hold. In fact, although often it is still possible to rewrite Eq.~(\ref{eq:diffusion}) in a form,
\begin{equation}
x_{k+1} = k(x_k) + \eta_k,
\end{equation}
in general $\thetab$ cannot be inferred from $k(x_k)$ \citep[see also ][]{tim00}. 
This has important consequences. For example, according to the frequency sampling, 
a given time series can display different types of nonlinearities or even appears as a linear process.
Moreover, in general,$\{ \eta_k \}$ are not independent, discrete Gaussian random quantities
even when $w(t)$ is Gaussian. 

These considerations indicate that discrete modelling suffers intrinsic limitations 
that make it unsuited to the description of physical systems. 
Consequently, an approach based on continuous models
is necessary. This requires the capability to estimate the parameters of a system of SDE from the observed time
series that, in its turn, requires tools that permits the numerical integration of this kind of equations.

\section{SDE for modelling time series} \label{sec:continuum}

\subsection{Numerical solution of SDE}

Although the theory behind SDE is complex, the numerical integration of this kind 
of equation does not present much more problems than the integration of the deterministic differential 
equations \citep{klo97, hig01}. Indeed, for a general SDE, 
\begin{equation}
\dot x(t) = a[t,x(t)] dt + b[t,x(t)] w(t),
\end{equation}
the simplest integration scheme, i.e. the Euler method, is given by
\begin{equation} \label{eq:euler}
x_{t_{k+1}} = x_{t_k} + \mu_{t_k} \Delta t_{k} + \sigma_{t_k} w_{t_{k}}.
\end{equation}
Here, $\mu_{t_k} = \mu[t_k,x_{t_k}]$, $\sigma_{t_k} = \sigma[t_k,x_{t_k}]$, $\Delta t_k = t_{k+1} - t_k$, 
and $\{ t_k \}_{k=0}^N$ represents a set of not necessarily equispaced time instants.
A pitfall of this method is that its order of strong convergence $\gamma$ is rather small, say $0.5$.
For this reason, for the solution of the SDE's in Figs.~ \ref{fig:series}, \ref{fig:pair}b
we have used the Milstein scheme 
\begin{equation} \label{eq:milst}
x_{t_{k+1}} = x_{t_k} + \bar \mu_{t_k} \Delta t_{k}
+ \sigma_{t_k} w_{t_{k}} \sqrt{\Delta t_{k}} + \frac{1}{2}~\sigma_{t_k}  \sigma'_{t_k} w_{t_k}^2 \Delta t_{k},
\end{equation}
that has $\gamma=1$. Here,
$\sigma_{t_k} = \sigma[t_k,x_{t_k}]$, $\bar \mu_{t_k} = \mu[t_k,x_{t_k}] - \frac{1}{2} \sigma_{t_k} \sigma'_{t_k}$,
and the symbol '' ${}'$ '' denotes differentiation with respect to $x(t)$. 
Integration schemes with higher values of $\gamma$ are possible, however they are rather more cumbersome to implement 
\citep[e.g., see ][]{klo97}. 
The algorithms~(\ref{eq:euler}) and ~(\ref{eq:milst}) can be easily
extended to deal with systems of SDE \citep{klo97}.

\subsection{Parameter estimation in SDE} \label{sec:sde}

In the past various approaches have been suggested to estimate the
parameters of SDE given discrete observations 
\citep[e.g., see ][ and reference therein]{bib95,tim00,bra02,dur02}. 
However, most of them have serious limits in the computational burden, and/or the
impossibility to deal with measurement errors, and/or the difficulty in the numerical
implementation.

Here, we present a general and flexible approach that is applicable to systems in the form
\begin{align}
\xdb(t) & = \fb\left[\xb(t), \uub(t), t ,\thetab \right] + \sigmab\left[\uub(t),t,\thetab\right] \wb(t); 
\label{eq:modelc1} \\
\yb_{t_k} & = \hb \left[\xb(t_k), \uub(t_k), t_k, \thetab \right] + \eb_{t_k},
\label{eq:modelc1b} 
\end{align}
where $\xb(t) \in \RB^n$ is the vector of the state variables, $\uub(t) \in \RB^m$ is the vector
of known input deterministic variables, $\yb_{t_k} \in \RB^l$ is the vector of observable variables, 
$\thetab \in \RB^p$
is the vector of parameters, $\fb[\cdot] \in \RB^n$, $\sigmab[\cdot] \in \RB^{n \times n}$
and $\hb[\cdot] \in \RB^l$ are (possibly) nonlinear functions, $\wb(t)$ is a n-dimensional,
standard, continuous white noise process with covariance given by the identity matrix, and $\eb_{t_k}$ 
is the vector of the measurements errors supposed
to be zero mean, Gaussian quantities with covariance matrix $\Sb(\uub_{t_k}, t_k, \thetab)$. The quantitites
$\{ \eb_{t_k} \}$  and $\wb(t)$ are supposed to be mutually independent for all $t$ and $t_k$.
This model is not so general as model~(\ref{eq:evolves1}). However, it is of interest 
in various practical applications. Moreover, as shown in Appendix~\ref{sec:appendixa}, through 
an appropriate transformation it is often 
possible to transform the more general state model
\begin{equation}
\xdb(t) = \fb\left[\xb(t), \uub(t), t ,\thetab \right] + \sigmab\left[\xb(t),\uub(t),t,\thetab\right] \wb(t), 
\end{equation}
(i.e., with $\sigmab[\cdot]$ that depends also on $\xb(t)$) to the form~(\ref{eq:modelc1}).

Given a particular model structure, the {\it maximum likelihood} (ML) estimation of the unknown parameters
can be performed by finding the parameters $\thetab$ that maximize the likelihood function of a given sequence of
measurements, say $\yb_{t_0}, \yb_{t_1}, \ldots, \yb_{t_k}, \ldots, \yb_{t_N}$. By introducing the notation
\begin{equation}
\Ym_k = [\yb_{t_k}, \yb_{t_{k-1}}, \ldots, \yb_{t_1}, \yb_{t_0}],
\end{equation}
the likelihood function is the joint probability density
\begin{equation}
L(\thetab; \Ym_N) = p(\Ym_N | \thetab),
\end{equation}
or equivalently:
\begin{equation} \label{eq:likelihood}
L(\theta; \Ym_N) = p(\yb_{t_0} | \thetab) \prod_{k=1}^N p(\yb_k | \Ym_{k-1}, \thetab).
\end{equation}
Here the rule $p(A \cap B) = p(A | B) p(B)$ has been applied to form a product of conditional probability
densities. In order to obtain an exact evaluation of the likelihood function, the initial probability density 
$p(\yb_{t_0} | \thetab)$ must be known and all the subsequent conditional densities must be determined by successively
solving Kolgomorov's forward equation and applying Bayes' rule \citep{jaz70}, but this approach is
computationally infeasible in practice. Given that in Eq.~(\ref{eq:modelc1}) the term containing
$\sigmab(t)$ does not depend on $\xb(t)$, a more efficient alternative can be proposed. In particular,
since the dynamics of model~(\ref{eq:modelc1}) is driven by Gaussian, white noise processes,
it is reasonable to assume that, under some regularity conditions, the conditional PDFs $p(\yb_k | \Ym_{k-1},\thetab)$
can be well approximated by Gaussians. Since the Gaussian density is completely characterized by its mean and 
covariance, by introducing the notation
\begin{align}
\widehat \yb_{t_k|t_{k-1}} & = \rm{E}\left\{\yb_{t_k} | \Ym_{k-1}, \thetab\right\}, \\
\Rb_{t_k|t_{k-1}} & = {\rm V}\left\{\yb_{t_k} | \Ym_{k-1}, \thetab\right\}, \\
\epsilonb_{t_k} & = \yb_{t_k} - \yhb_{t_k|t_{k-1}},
\end{align}
where ${\rm E}[\cdot]$ and ${\rm V}[\cdot]$ are, respectively, the the mean and covariance operators,
the likelihood function~(\ref{eq:likelihood}) can be written in the form
\begin{equation}
L(\thetab; \Ym_N) = p(\yb_{t_0} | \thetab) \prod_{k=1}^N \frac{\exp\left(- \frac{1}{2} 
~\epsilonb_{t_k}^T \Rb^{-1}_{t_k| t_{k-1}} \epsilonb_{t_k} \right)}{\left(\det[\Rb_{t_k|t_{k-1}}]\right)^{1/2} 
\left(2 \pi \right)^{l/2}}.
\end{equation}
For a fixed $\thetab$, the quantities $\epsilonb_{t_k}$ and $\Rb_{t_k|t_{k-1}}$ can be computed
by means of an {\it extended Kalman filter} (see Appendix~\ref{sec:appendixb}). Further, conditioning on $\yb_{t_0}$ 
and taking the 
negative logarithm $\l(\thetab) = - \ln\left[ L(\thetab; \Ym_k | \yb_{t_0}) \right]$ gives
\begin{equation}
l(\thetab) \propto
\sum_{k=1}^N \left(\ln(\det[ 
\Rb_{t_k|t_{k-1}}]) +\epsilonb_k^T \Rb^{-1}_{t_k|t_{k-1}} \epsilonb_{t_k} \right).
\end{equation}
The ML estimate of $\thetab$ (and optionally of $\yb_{t_0}$) can be now determined by solving the
nonlinear optimization problem
\begin{equation} \label{eq:thetah}
\thetawb = \underset{\thetab}{\arg\min} \left[ l(\thetab) \right].
\end{equation}
An estimate of the uncertainty of $\thetawb$ is obtained by the fact that the ML-estimator is asymptotically normal
with mean $\thetab$ and covariance $\Sigmab$ given by the lower bound of the Cramer-Rao inequality, i.e.,
\begin{equation}
\Sigmab = \Hb^{-1}.
\end{equation}
Here, the Hessian matrix $\Hb = \{ h_{ij} \}$ is given by
\begin{equation}
h_{ij} = - {\rm E} \left\{ \frac{\partial^2 l(\thetab)}{\partial \theta_i \partial \theta_j} \right\},
\end{equation}
that can be estimated with
\begin{equation}
h_{ij} \approx - \left\{ \frac{\partial^2 l(\thetab)}{\partial \theta_i \partial \theta_j} \right\}_{\thetab = \thetawb}.
\end{equation}

\section{A worked example} \label{sec:worked}

\begin{figure}
        \resizebox{\hsize}{!}{\includegraphics{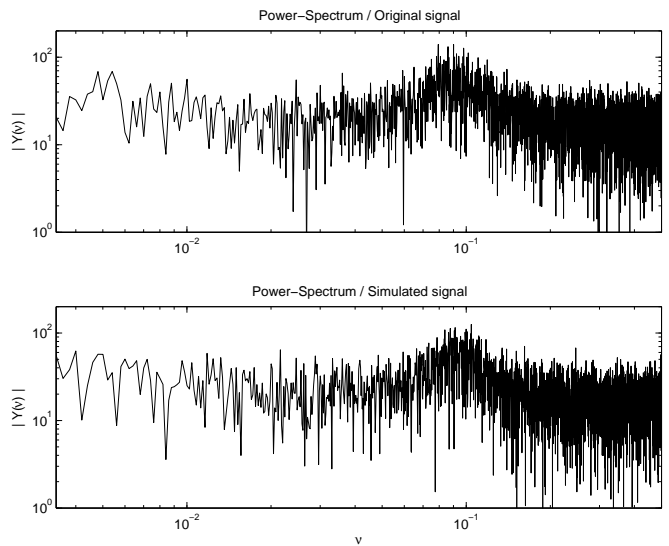}}
        \caption{Upper panel: power-spectrum $| Y(\nu) |$ of the first $5000$ data of the time series of SCO-X1;
        Lower panel: power-spectrum $| Y(\nu) |$ of a typical $5000$ points realization of the process 
        given by the fitted model~(\ref{eq:sim1})-(\ref{eq:sim2}). The frequency $\nu$ is in Nyquist units.}
        \label{fig:fig_sim2}
\end{figure}

To illustrate the effective potentialities of stochastic modelling in the analysis of
time series, we have considered a sequence of X-ray observations 
of low mass X-ray binary Sco X-1 (van der Klis, priv. comm.) made with the 
{\it Proportional Counter Array} (PCA) on the Rossi-XTE spacecraft \citep{bra93}. The time series used
in the experiment contains $1000$ data. Sampling is regular with a time step of 0.015 seconds. As shown
in Fig.\ref{fig:fig_sim2}, the time series of this object 
presents a power-spectrum typical of the quasi-periodic objects (QPO), i.e.,
a broad peak superimposed to a steeply decreasing continuum. If one interprets such a fact as due to a single 
driving mechanism, a possible model for reproducing the observed signal is
\begin{align}
\dot x(t) & = - a x(t) + \sigma_1 w(t); \label{eq:sim1} \\
y_k & = x(t_k) + A \cos[ \omega_0 t_k + \sigma_2 x(t_k)] + e_k.   \label{eq:sim2}
\end{align}
Here, $a$, $A$, $\sigma_1$, and $\sigma_2$ are unknown constants that are to be estimated. 
The frequency $\omega_0$ and the variance of
the measurement errors $e_k$ (assumed i.i.d Gaussian) are estimated through the central position of the
broad peak and the high frequency level in the estimated power-spectrum, respectively.
The process $x(t)$ is assumed to be not observable. 

\begin{figure}
        \resizebox{\hsize}{!}{\includegraphics{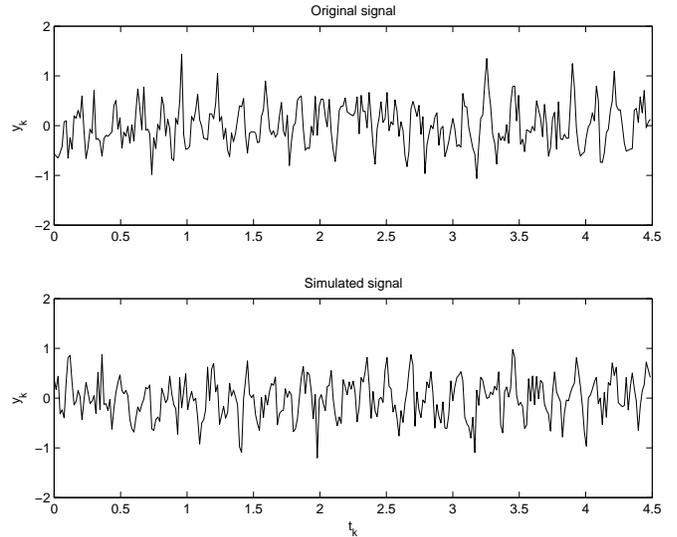}}
        \caption{Upper panel: original (mean subtracted) time series of SCO-X1;
        Lower panel: typical time series obtainable through the fit of 
        model~(\ref{eq:sim1})-(\ref{eq:sim2}) (see text). Although the time series used in the analysis contains
        $1000$ data, 
        here, for easiness of comparison, only the first $300$ data are shown. Time is in unit of {\it second}.}
        \label{fig:fig_sim1}
\end{figure}

The idea behind this model is that the luminosity of the object is determined by a driving linear
stochastic process $x(t)$ (e.g., the accretion rate) superimposed to a periodic process that, however,
is perturbed by $x(t)$ itself. Figures~\ref{fig:fig_sim1} and \ref{fig:fig_sim2} show that the results obtainable 
with the methodology explained in the previous section are remarkably good. Of course, this does not mean
that such a simple model corresponds to a real scenario. Actually, the only thing that it is possible to claim is the
compatibility of model~(\ref{eq:sim1})-(\ref{eq:sim2}) with the observations. However, we stress 
that this is the most it can be obtained from any technique of data analysis.


\section{Conclusions} \label{sec:conclusions}

The time series usually available in astronomy are able to characterize only a subset of the system of equations that describe the dynamics of the physical system under study. For this reason, although in principle it is always possible to determine a statistical model able to reproduce the experimental data, without any a priori physical model there are not so many possibilities to obtain a reliable reconstruction of the physical scenario investigated. In general, this means that an approach to the analysis of time series exclusively based on the experimental data will provide inconclusive results,
and that the practice to search for more and more sophisticated statistical techniques is not very productive. In many situations, the only possibility to get some physical insights is to carry out the analysis in a well defined physical context. In this respect, {\it stochastic dynamics}, i.e. modelling the time series with stochastic differential equations, appears to be a promising tool. A benefit of working within such a framework is that one is forced to provide a mathematical/physical formalization of the starting hypotheses adopted in the analysis of the signals (e.g. linearity, non-linearity, type of non-linearity, ...). In this way, there is no risk of misunderstandings concerning the interpretations of the final results. Additionally, a more direct relation between complex physical models and the limited observational material is made possible permitting a more efficient interaction between data and theory. 

A code, implementing some efficient numerical tools for modelling the experimental time series with stochastic 
differential equations, is freely downloadable from {\it http://www.imm.dtu.dk/ctsm/}.

\begin{acknowledgements}
We thank Prof. M. van der Klis for the availability of the unpublished data on Sco X-1.
\end{acknowledgements}

\appendix
\section{A multivariate transformation} \label{sec:appendixa}

In this section, a bijective transformation 
\begin{equation} \label{eq:transformation}
\zb(t) = \Psib[\xb(t),t]
\end{equation}
is proposed to transform the state-equation
\begin{equation} \label{eq:original}
\xdb(t) = \fb[\xb(t), \uub(t), t ,\thetab ] + \sigmab[\xb(t),\uub(t),t,\thetab ] \wb(t), 
\end{equation}
to
\begin{equation} \label{eq:transformed}
\zdb(t) = \ftb[ \zb(t), \uub(t), t ,\thetab ] + \sigmatb[\uub(t),t,\thetab ] \wb(t).
\end{equation}
Here, $\Psib[\cdot]$ is assumed continuously differentiable with respect to $t$ and twice continuously differentiable with respect to $\xb(t)$. The covariance matrix of $\wb(t)$ is assumed to be equal to the identity matrix.
Further assumptions are:
\begin{enumerate}
\item all the elements in $\sigmab [\xb(t),\uub(t),t,\thetab ]$ are strictly nonzero, i.e. 
\begin{equation}
\sigma_{ij}[\xb(t), \uub(t), t, \thetab] \neq 0, \qquad i,j = 1, 2, \ldots, n;
\end{equation}
\item For each $i$ there exists only one $\sigma_{ij}$ as a function of one and only one state-variable $x_{\nu(i)}(t)$,
where $\nu(i)$ is different for each $i$, i.e.,
\begin{equation}
\sigma_{ij}[\xb(t),\uub(t),t,\thetab ] = \sigma_{ij}[x_{\nu(i)}(t),\uub(t),t,\thetab ];
\end{equation}
\item The functions $\sigma_{ij}[x_{\nu(i)}(t),\uub(t),t,\thetab ]$ are bijective and 
$\sigma_{ij}^{-1}[x,\uub(t),t,\thetab ]$ are integrable with respect to
$x$. 
\end{enumerate}
Given these assumptions, it can be shown \citep{nie01} that the transformation
\begin{equation}
\Psi_l [x_{\nu(i)}(t)] = \int \frac{dx}{\sigma_{ij}[\xb(t), \uub(t), \thetab]} \Big\vert_{x = x_{\nu(i)}(t)},
\end{equation}
$l,i,j = 1,2,\ldots,n$, fullfills Eq.~(\ref{eq:transformed}).

It is interesting to notice that, after the transformation~(\ref{eq:transformation}), the 
state-equation~(\ref{eq:transformed}) contains the same parameters $\thetab$ as the original 
state-equation~(\ref{eq:original}). Moreover, the measurement-equation~(\ref{eq:modelc1b}), 
\begin{equation}
\yb_{t_k} = \hb \left[\xb(t_k), \uub(t_k), t_k, \thetab \right] + \eb_{t_k}, \nonumber
\end{equation}
can be used in its original form since the state-equation $\xb(t)$ is obtainable
from inverse transformation $\xb(t) = \Psib^{-1}[\zb(t),t]$.

\section{Kalman Filter for estimating parameters in SDE} \label{sec:appendixb}

In Sec.~\ref{sec:sde} it is shown that the parameters $\thetab$ of a system of 
SDEs can be estimated through the maximization of the likelihood function~(\ref{eq:thetah}). 
This requires the quantities $\epsilonb_{t_k}$ and $\Rb_{t_k,t_{k-1}}$ that, however, are unknown and have to
be estimated. Here, we propose an approach based on the {\it continuous-discrete extended Kalman filter}.
In particular, if at a given time instant $t_k$ the quantities $\thetab$, 
$\xhb_{t_k | t_{k-1}} = {\rm E}[\xb_{t_k} | \xb_{t_{k-1}}]$, 
and $\PPb_{t_k | t_{k-1}} = {\rm E}[\xb_{t_k} \xb_{t_k}^T | \xb_{t_{k-1}}]$
are fixed, then the procedure is based on the iterated solution of the following sequence of equations
\begin{enumerate}
\item -  The {\it output prediction equations}
\begin{align}
\yhb_{t_k | t_{k-1}} & = \hb(\xhb_{t_k | t_{k-1}}, \uub_{t_k}, t_k, \thetab); \\
\Rb_{t_k | t_{k-1}} & = \Cb \PPb_{t_k | t_{k-1}} \Cb^T + \Sb;
\end{align} 
\item - The {\it innovation equation}
\begin{equation}
\epsilonb_{t_k} = \yb_{t_k} - \yhb_{t_k | t_{k-1}};
\end{equation}
\item - The {\it Kalman gain equation}
\begin{equation}
\Kb_{t_k} = \PPb_{t_k | t_{k-1}} \Cb^T \Rb^{-1}_{t_k | t_{k-1}};
\end{equation}
\item The {\it updating equations}
\begin{align}
\xhb_{t_k | t_k} & = \xhb_{t_k | t_{k-1}} + \Kb_{t_k} \epsilonb_{t_k}; \\
\PPb_{t_k | t_k} & = \PPb_{t_k | t_{k-1}} - \Kb_{t_k} \Rb_{t_k | t_{k-1}} \Kb_{t_k}^T;
\end{align}
\item - The {\it prediction equations}
\begin{align}
\frac{d \xhb_{t|t_k}}{dt} & = \fb(\xhb_{t|t_k}, \uub_t, t, \thetab), \quad t \in [t_k, t_{k+1});  \label{eq:eq1} \\
\frac{d\PPb_{t|t_k}}{dt} & = \Ab \PPb_{t|t_k} + \PPb_{t|t_k} \Ab^T + \sigmab \sigmab^T, \quad t \in [t_k, t_{k+1}).
\label{eq:eq2}
\end{align}
\end{enumerate}
Equations~(\ref{eq:eq1}) and (\ref{eq:eq2}) provide the quantities $\xhb_{t_{k+1} | t_k}$ and $\PPb_{t_{k+1} | t_k}$ 
that can be used to start a new iteration of the sequence.
Here, $\sigmab=\sigmab(\uub_{t_k}, t_k, \thetab)$, $\Sb=\Sb(\uub_{t_k}, t_k, \thetab)$, and
\begin{align}
\Ab & = \frac{\partial \fb}{\partial \xb_t} \Bigg\vert_{\xb=\xhb_{t_k | t_{k-1}, \uub_{t_k}, t=t_k, \thetab}}, \\
\Cb & = \frac{\partial \hb}{\partial \xb_t} \Bigg\vert_{\xb=\xhb_{t_k | t_{k-1}, \uub_{t_k}, t=t_k, \thetab}}.
\end{align}
Initial conditions
for the iteration are $\xhb_{t|t_0} = \xb_{t_0}$ and $\PPb_{t|t_0} = \PPb_{t_0}$, which may either be 
pre-specified or considered as additional parameters to estimate.

A pitfall of the above procedure is that the matrices $\Ab$ and $\Cb$ have been obtained through the
linearization of the functions $\fb(\cdot)$ and $\hb(\cdot)$. Therefore, the approximated solutions obtained
by solving Eqs.~(\ref{eq:eq1}) and (\ref{eq:eq2}) may bee too crude. Moreover, the assumption of
Gaussian conditional densities is only likely to hold for small sample times. To alleviate these problems,
a better approximation is obtainable through a subsampling of
the time interval $[t_k, t_{k+1})$, i.e., $[t_k, \ldots, t_j, \ldots, t_{k+1})$, and the linearization of
Eqs.~(\ref{eq:eq1}), (\ref{eq:eq2}) at each of such subsampling instants. This also means that 
the direct numerical solution
of Eqs.~(\ref{eq:eq1}), (\ref{eq:eq2}) can be avoided by applying the analytical solutions to the
corresponding linearized prediction equations 
\begin{align}
\frac{d \xhb_{t|t_j}}{dt} & = \fb(\xhb_{t|t_j}, \uub_{t_j}, t_j, \thetab) + \Ab(\xhb_t - \xb_{t_j})
+\Bb(\uub_t - \uub_{t_j}), \label{eq:eq1a} \\
\frac{d\PPb_{t|t_j}}{dt} & = \Ab \PPb_{t|t_j} + \PPb_{t|t_j} \Ab^T + \sigmab \sigmab^T, \label{eq:eq2a}
\end{align}
where $t \in [t_j, t_{j+1})$, $\sigmab=\sigmab(\uub_{t_j}, t_j, \thetab)$, $\Sb=\Sb(\uub_{t_j}, t_j, \thetab)$, and
\begin{align}
\Ab & = \frac{\partial \fb}{\partial \xb_t} \Bigg\vert_{\xb=\xhb_{t_j | t_{j-1}}, \uub_{t_j}, \thetab}, \\
\Bb & = \frac{\partial \fb}{\partial \uub_t} \Bigg\vert_{\xb=\xhb_{t_j | t_{j-1}}, \uub_{t_j}, \thetab}.
\end{align}
The solution
of Eq.~(\ref{eq:eq2a}) is given by
\begin{equation}
\PPb_{t_{j+1} | t_j } = \Phib_s \PPb_{t_j | t_j} \Phib_s^T + \int_{0}^{\tau_s} \Phib_s
\sigmab \sigmab^T \Phib_s^T ds,
\end{equation}
where $\tau_s = t_{j+1} - t_j$, and $\Phib_s = {\rm e}^{\Ab s}$. The solution of Eq.~(\ref{eq:eq1a}) is more
difficult to find. If $\Ab$ is nonsingular, it is given by
\begin{equation}
\xhb_{t_{j+1} | t_j}
= \xhb_{t_j | t_j} - \tau_s \Ab^{-1} \Bb \alphab + \Ab^{-1} (\Phib_s - \Ib)
(\Ab^{-1} \Bb \alphab + \fb),
\end{equation}
where
\begin{equation}
\alphab = \frac{\uub_{j+1} - \uub_j}{t_{j+1} - t_j},
\end{equation}
and $\fb = \fb(\xhb_{t|t_j}, \uub_{t_j}, t_j, \thetab)$.
Things are more complex if $\Ab$ is singular. More details can be found in \citet{kri04}.

\end{document}